\documentclass[%
 reprint,
 longbibliography,
 amsmath,
 amssymb,
 aps,
 pre,
]{revtex4-1}

\usepackage{dcolumn}
\usepackage{bm}

\usepackage{amssymb}
\usepackage{amsmath}
\usepackage{cases}
\usepackage{graphicx,color}
\usepackage{array}
\usepackage{multirow}
\usepackage{makecell}
\usepackage{mathtools}

\definecolor{r}{rgb}{1,0,0}   
\definecolor{g}{rgb}{0,1,0}   
\definecolor{b}{rgb}{0,0,1}
\definecolor{purple}{rgb}{0.808,0.454,0.718}

\begin{document}


\title{Quantifying the Long-Range Structure of Foams and Other Cellular Patterns with Hyperuniformity Disorder Length Spectroscopy}


\author{A. T. Chieco \& D. J. Durian}
\affiliation{Department of Physics and Astronomy, University of Pennsylvania, Philadelphia, PA 19104-6396, USA}


\date{\today}

\begin{abstract}
We investigate the local- and long-range structure of four different space-filling cellular patterns: bubbles in a quasi-2d foam plus Voronoi constructions made around points that are uncorrelated (Poisson patterns), low discrepancy (Halton patterns), and displaced from a lattice by Gaussian noise (Einstein patterns). We study distributions of local quantities including cell areas and topological features; the former is the widest for bubbles in a foam making them locally the most disordered but the latter show no significant differences between the cellular patterns. Long-range structure is probed by the spectral density and also by converting the real-space spectrum of number density or volume fraction fluctuations for windows of diameter $D$ to the effective distance $h(D)$ from the window boundary where these fluctuations occur. This real-space hyperuniformity disorder length spectroscopy is performed on various point patterns which are determined by the centroids of the bubbles in the foam, by the points patterns around which the Voronoi cells are created and by the centroids of the Voronoi cells. These patterns are either unweighted or weighted by the area of the cells they occupy. The unweighted bubble centroids have $h(D)$ that collapses for the different ages of of the foam with random Poissonian fluctuations at long distances. All patterns of area-weighted points have constant $h(D)=h_e$ for large $D$; $h_e=0.084 \sqrt{\left< a\right>}$ for the bubble centroids is the smallest value, meaning they are most uniform. All the weighted centroids collapse to the same constant $h_e=0.084 \sqrt{\left< a\right>}$ as for the foams. A similar analysis is performed on the edges of the cells where the spectra of $h(D)$ for the foam edges show $h(D) \sim D^{1-\epsilon}$ where $\epsilon=0.3$.
\end{abstract}



\maketitle




\section{Introduction}
%
%
%
%

There are many well-established ways to quantify the local structure of foams and other cellular systems using cell size, shape, topology, and neighbor correlations \cite{WeaireHutzlerBook,GlazierWeaireRev,StavansRev,FlyvbjergPRE}. Distributions of such local measures are used to describe the entire foam packing and under proper normalization they remain the same as the foam coarsens; {\it i.e.}\ they exhibit statistical self-similarity \cite{StavansPRA, StavansPhysicaA, StavansGlazierPRL, deIcazaJAP76, GlazieretalPMB, HerdtleArefJFM, SegelEtAlPRE, RutenbergMcCurdyPRE, NeubertSchreckPhysicaA}. By contrast, quantifying the long range structure of foams and other cellular systems remains an open question.

Recently the concept of hyperuniformity was introduced regarding the structure in disordered materials at long distances \cite{TorquatoPRE2003, ZacharyJSM2009, TorquatoReview}. Materials are called ``hyperuniform" if long range density fluctuations are suppressed to the same extent as in crystals. Work done on hard-particle packings of bidisperse disks, ellipses and superballs at the jamming transition found they are hyperuniform and the researchers posit that hyperuniformity exists in all systems at the jamming transition regardless of particle shape or polydispersity \cite{ZacharyPRL2011, ZacharyPRE2011_Disks, ZacharyPRE2011_Ellipse}.  Analysis on a wide assortment of other disordered materials at or slightly below the jamming transition finds they have signatures of hyperuniformity \cite{KuritaPRE2011, BerthierPRL2011, JiaoPRE2014, DreyfusPRE2015, WeijsPRL2015, AtkinsonPRE2016}. However, hyperuniformity is not a signature of all jammed systems and work on simulated packing of bidisperse soft disks demonstrates it does not exist above the jamming transition \cite{WuPRE2015, IkedaPRE2015, ATCjam}. Additionally Ref.~\cite{ATCjam}  shows for 2-dimensions and Ref.~\cite{IkedaPRE2015} shows for 3-dimensions that not only does hyperuniformity not exist above jamming but the overall uniformity of the packing decreases as the distance above jamming increases. This is where foams pose an interesting problem. Foam is far above the jamming transition where hyperuniformity has not been observed, yet it is completely space-filling and the total absence of density fluctuations makes it trivially hyperuniform. Nevertheless the bubble packing structure and the distribution of liquid films are visually disordered, possessing large spatial fluctuations that could impact behavior and need to be quantified.

While foams are a naturally occurring cellular solid, this same problem exists for any disordered system with global packing fraction $\phi=1$.  To examine this problem in detail we generate space-filling cellular packings by partitioning space with Voronoi constructions around point patterns of varying disorder. Such patterns were analyzed in recent studies with regards to their long range uniformity \cite{KlattUniversal2019,KimCellScaling2019}. In Ref.~\cite{KimCellScaling2019} the authors, using a usual method for diagnosing hyperuniformity, find that for small-$q$ wave vectors the scaling of the spectral density behaves like $\chi(q) \sim q^{4}$ where the exponent is exact based on the conditions of their simulation. They do not present experimental data, so we perform the same kind of Fourier analysis as they do for our foam systems as well as our Voronoi constructions. Since all of the packings closely mirror the conditions analyzed in Ref.~\cite{KimCellScaling2019}, we are interested in whether analyzing our systems recover the same spectral density scaling exponent. This also allows us to test the extent to which foams are hyperuniform and more generally we compare the uniformity of their long range structure to the other space filling cellular patterns.

In real space the method to test for hyperuniformity is to randomly placing a series of local observation windows throughout a sample, measuring the area fraction covered by the particles that land within each window and calculating the variance for the set of measured area fractions; this is repeated for growing observation windows, and if at large length scales the variance is suppressed to the same extent as in crystals then the system is said to be hyperuniform. There are two ways to define the area fraction within a measuring window. One method calculates the area covered by a particular phase of the media that lands inside the window. If a cellular packing has global packing fraction $\phi=1$ then the local area fractions are $\phi_w=1$ for every measuring window and no meaningful signature of hyperuniformity can be found. The other method, which we employ here, defines cells as a point weighted by the area of the cell; if that point lands inside the local observation window then the entire area of the cell is counted but if it lands outside the observation window then none of the area is considered.

Measuring the asymptotic scaling behavior provides an answer to whether these systems are hyperuniform but does not provide additional insight into the actual structure of the underlying pattern. This can be done in principle using the same tools used to diagnose hyperuniformity but a necessary step is converting the fluctuations observed in a local measurement window into a length scale; this length scale is called the hyperuniformity-disorder length $h$ and its size is the distance from the boundary of a local measurement window where particles set number density fluctuations \cite{ATCpixel,DJDhudls}. Therefore the value of $h$ provides us with a length scale for disorder that probes the nature of long-range structure as well as the structure at smaller distances. This technique is called ``hyperuniformity disorder length spectroscopy" (HUDLS) and has shown success in identifying long range structure for other soft systems \cite{ATCjam}. Here we use it to uncover and compare the extent of potential hidden order of various structural features of foam and other cellular patterns. We are also able to determine whether the local structure informs long range structure. 

\section{Methods}

\subsection{Hyperuniformity: Scaling and Definitions} 

In this section we begin with an optional review of established methods we shall use to diagnose hyperuniformity in ways that quantify long-range structure.  This may be done either by the asymptotic scaling of either the spectral density $\chi(q)$ or by the variance $\sigma_\phi^2(D)$ in the set of local volume fractions measured in randomly-placed windows of diameter $D$. If the spectral density has small wave vector behavior like $\chi(q)\sim q^\epsilon$ with $\epsilon>0$, or more generally if $\chi(0^+)=0$, then a system is said to be hyperuniform.  Scaling with $0 < \epsilon \leq 1$ corresponds to the long length scaling $\sigma_\phi^2(D)\sim 1/D^{d+\epsilon}$ where $d$ is dimensionality; for $\epsilon \ge 1$, $\chi(q)\sim q^\epsilon$ corresponds to $\sigma_\phi^2(D)\sim 1/D^{d+1}$ and we say the system is strongly hyperuniform. By contrast, ordinary systems exhibit Poissonian fluctuations where $\epsilon=0$. In reciprocal space the spectral density is $\chi(0^+)=C$ where $C>0$ is some constant and the volume fraction variance scales like $\sigma_\phi^2(D)\sim 1/D^d$ according to the dimensionality. The power $\epsilon$ can be used as a proxy for order, but the actual value does not have an intuitive physical interpretation. 

It is important to have meaningful interpretations of order from the actual data. For the spectral density, this is achieved by choosing a proper normalization of $\chi(q)$ by comparing the data to the that for a ``Poisson pattern" where particles are placed totally at random. If a cellular patterns in 2-dimensions is represented by a central-point with the entire area $a_j$ of particle $j$ is at the location ${\bf r}_j$ of its center then a suitable definition of the spectral density is
\begin{equation}
	\chi(q) \equiv {  \left(  \sum a_j e^{i {\bf q}\cdot{\bf r}_j}~\sum a_k e^{-i {\bf q}\cdot{\bf r}_k}\right) }/{\sum a_j^2 }
\label{chidef}
\end{equation}
where $q=|{\bf q}|$ for isotropic packings and the sums are over all particles. This normalization means Poisson patterns have $\chi(q)=1$ which becomes a nominal upper bound and insight into structure at a given $q$ is extracted from how far $\chi(q)$ lies below this value. Another benefit of this normalization is for systems of monodisperse particles or for point patterns the spectral density reduces to the structure factor, $S\left(q\right)$.  

In real space, order is determined from the spectrum of hyperuniformity disorder lengths $h(D)$. Determining $h(D)$ for 2-dimensional systems first requires  finding the area fraction variance  $\sigma_\phi^2(D)$. To find $\sigma_\phi^2(D)$ the variance is measured from a set of local area fractions $\sum N_i a_i /A_\Omega$, where $N_i$ is the number of particles of species $i$ whose centers lie inside a randomly placed window of area $A_\Omega=\pi (D/2)^2$ and the sum is over species. This is the real space definition of the central point representation. Using these definitions a completely random arrangement of particles will have $\sigma_\phi^2\left(D\right) = \left< a \right> / A_{\Omega}$ where $\langle a \rangle = \sum \phi_i {a_i}/ \sum \phi_i$ is the area fraction weighted average particle area, $\phi_i$ is the area fraction covered by particles with $a_i$, and $\phi=\sum \phi_i$ is the area fraction of all the particles in the system.

\begin{figure}
\includegraphics[width=2.2in]{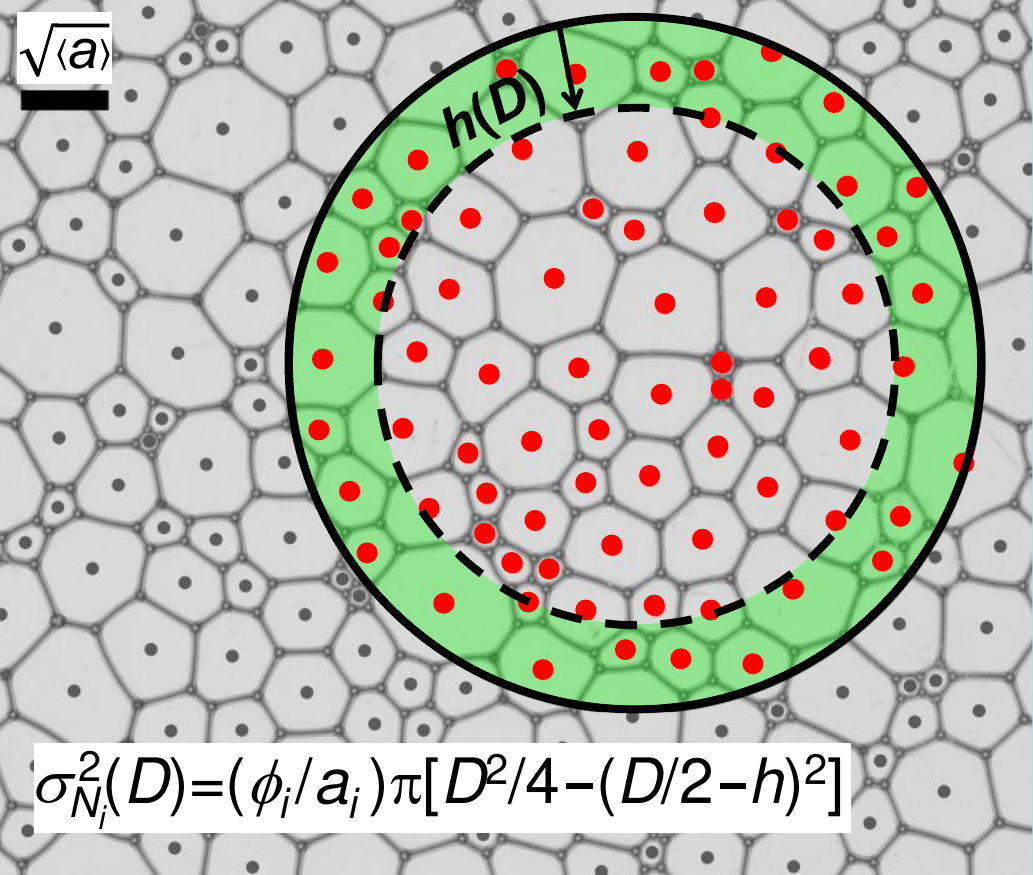}
\caption{Image of a quasi 2-dimensional foam with the bubble centroids marked by dots. The total area of bubbles enclosed in a circular window is taken as the sum of areas for bubbles with enclosed centroids (red).  The area fraction variance is controlled by the number of particles in the shaded region of thickness $h(D)$, averaged over window placements.  As depicted here for $D=8 \sqrt{\left<a\right>}$, the hyperuniformity disorder length is $h(D)=\sqrt{\left<a\right>}$ where $\langle a\rangle$ is the area-fraction weighted average bubble area. The value of $h(D)$ is inflated by over $10\times$ its actual value for illustrative purposes.}
\label{hdefpic}
\end{figure}

The measured area fractions fluctuate depending on where the measuring window lands within the system; hyperuniform configurations have fluctuations that are understood to be due to particles at the surface of the measuring windows \cite{TorquatoPRE2003}. Since particle centers do not actually lie {\it on} the window surface, it is more appropriate to picture fluctuations as determined by the average number of particles whose centers lie {\it within some distance} $h$ of the surface. For circular windows with area $A_\Omega= \pi (D/2)^2$ we can thus define $h$ from the number variance via $\sigma_{N_i}^2 = (\phi_i / a_i) \pi [(D/2)^2 - (D/2-h)^2]$, which is shown pictorially in Fig.~\ref{hdefpic}. The number variance is converted to an area fraction variance $\sigma_\phi^2=\sum \sigma_{N_i}^2 a_i^2/A_\Omega^2$ which leads to the following explicit definition of $h(D)$ in terms of the measured variance:
\begin{eqnarray}
	\frac{ \sigma_\phi^2(D) }{\phi} &\equiv& \frac{\langle a\rangle}{\pi \left(D/2\right)^2 }\left\{ 1 - \left[1-\frac{h(D)}{D/2} \right]^2\right\} \label{hdef}, \\
						     &\approx& 2 \frac{ \langle a\rangle h(D) }{ \left(D/2\right)^3 }\ {\rm for}\ D\gg h(D). \label{hdefapprox}
\label{hdefeq}
\end{eqnarray}
Accordingly, smaller $h(D)$ means more uniformity, larger $h(D)$ means more disorder, and $\sigma_\phi^2(D)\sim1/D^{d+\epsilon}$ corresponds to $h(D)\sim D^{1-\epsilon}$.  Poissonian fluctuations have $\epsilon=0$ and correspond to $h(D) \sim D$; the upper bound is $h(D)=D/2$ for a Poisson pattern.  Strong hyperuniformity where $\epsilon \ge 1$ corresponds to a large-$D$ asymptote that is constant: $h(D)=h_e$.  For this case, $\sigma_\phi^2(D)\sim \langle a\rangle h_e /D^{3}$ is made dimensionally correct by the existence of $h_e$ as an emergent length rooted in the intuitive notion of what it means to be hyperuniform. Thus $h_e$ is the desired measure of structure that is independent of $D$ when the system is hyperuniform, and Eq.~(\ref{hdef}) generalizes upon this to systems with any degree of uniformity. 

The definitions for the hyperuniformity-disorder length are discussed in much more detail in \cite{ATCpixel, DJDhudls}. These references also go through the calculations of the upper bound $h(D)=D/2$ as well as a lower bound for the ``separated-particle limit" where the size of the measuring window is smaller than the average distance between two particles. Additional discussion about our methods calculating the spectral density can be found in the appendix of \cite{ATCjam}.

We also note that hyperuniformity is truly a measure of number fluctuations and because points are given a weight equal to their area the above discussion is in the context of long wavelength area fraction fluctuations. However, using the central point representation fluctuations in any order parameter can be determined by assigning an appropriate weight to each point  {\it i.e.}\ for fluctuations in coordination number each point is given a weight equal to its number of contacts \cite{HexnerLiuNagel2018}. Assigning equal weight to each point makes the system monodisperse and it is treated simply as a point pattern; the signature of hyperuniformity for these systems is fluctuations in the number variance that grow more slowly than the volume of the window. This treatment changes the definitions and bounds from above: the spectral density reduces to the structure factor $S(q)$ and all particle ``areas" are $a_j=1$; the random expectation for the number variance is $\sigma_N^2 \left(D\right)=\rho A_\Omega$ where $\rho$ is the number density. The definition of $h(D)$ also changes and is defined by rearranging $\sigma_{N}^2 = \rho \pi [(D/2)^2 - (D/2-h)^2]$; $h(D)$  whether defined from the number variance or the area fraction variance is calculated from a ratio of the measured variance to the expected variance for a totally random system in both cases and our intuition for what it measures remains the same. Because foams have not been studied in the context of hyperuniformity, we explore our systems both as monodisperse point patterns using the centroids of the bubbles and as polydisperse systems where the centroids are weighted by their bubble area.

\begin{figure*}
\includegraphics[width=7in]{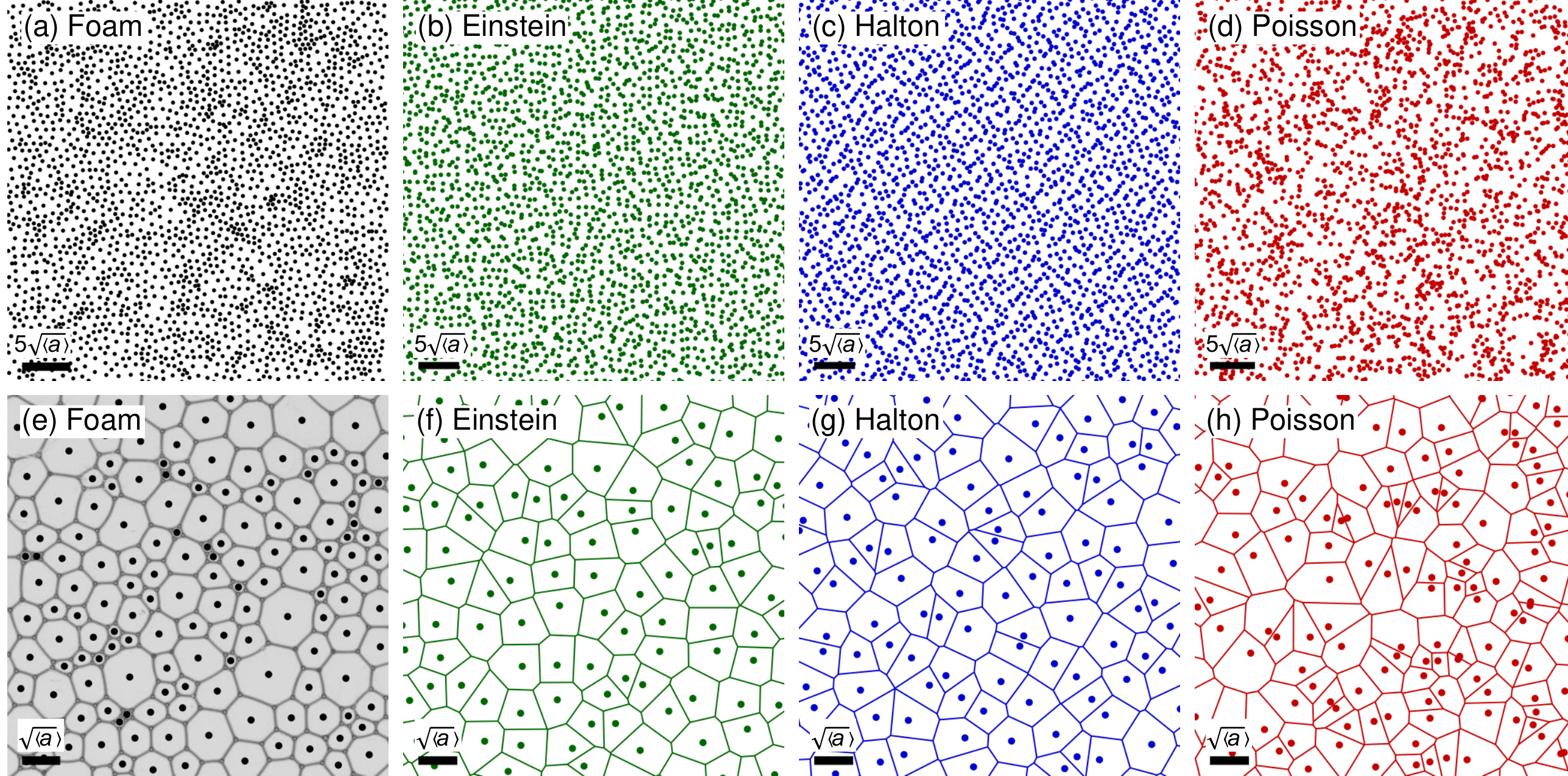}
\caption{Disordered points patterns: (a) shows the locations of the centroids of the bubbles in a quasi 2-dimensional foam; (b) the green dots are an Einstein pattern where points are randomly displaced from a square lattice with RMS displacement $\delta / b=0.26$ where $b$ is the lattice spacing; (c) the blue dots are a Halton set which is a low discrepancy pattern where points are determined algorithmically; (d) the dark red dots are a Poisson pattern where points are placed totally at random. Parts (e-h) show the cellular patterns used to partition space around the corresponding point pattern: Part (e) displays the bubbles of a quasi-2d foam as well as the bubble centroids; parts (f-h) show the cells of a Voronoi construction which are created around the points that occupy each cell. For analysis of area fraction fluctuations all points are given a value equal to the area of the cell they occupy.}
\label{PointRow}
\end{figure*}


\subsection{Foam and Voronoi Data} \label{methods}

We study foam made from a solution that is 75\% deionized water, 20\% glycerin and 5\% Dawn Ultra Concentrated Dish Detergent. It is generated inside a sample cell made from two 1.91~cm-thick acrylic plates separated by a spacing of 0.3~cm and sealed with two concentric o-rings, the inner of which has a 23~cm diameter; this is the same apparatus used in \cite{RothPRE2013} for foam coarsening experiments. Foams are produced as follows.  First the trough is filled with the desired amount of liquid, then flushed with Nitrogen and sealed. The entire sample cell is vigorously shaken for several minutes until the gas is uniformly dispersed as fine bubbles that are small compared to the gap between plates. The foam is thus initially very wet, opaque, and three-dimensional. The cell is immediately placed above a Vista Point~A light box and below a Nikon~D90 camera with a Nikkor AF-S 300mm 1:2.8D lens. After a few hours, the bubbles become large compared to the gap and the foam has coarsened into a quasi two dimensional state; once the foam is quasi-2d, images of it are taken every 2~minutes for 24~hours.

To gather relevant data for bubbles, such as their locations and areas, we first have to reconstruct the foam microstructure and film network. The reconstruction methods are described more thoroughly in the supplemental materials of \cite{ATCcoarsen}. More briefly, the first step is to locate the  vertices via  a convolution method of an example vertex structuring element and the foam  image. After the vertex locations are  identified they are connected to their neighbors by exploiting Plateau's laws.  Plateau's laws in 2-dimensions say that vertices are the junction of three films which meet at $120^\circ$ and that pairs of vertices are connected by films that are arcs of circles. Therefore we know where to look for neighboring vertices and once neighbors are identified we find equations for the circular arcs that connect them. Finally bubbles are identified by making closed loops of vertices. 

Analysis for hyperuniformity is ultimately performed on point patterns that represent the bubbles and the bubble centroids $(x_c,y_c)$ are a logical pattern to choose. These points are defined as
\begin{eqnarray}
	x_c &=&\sum{\left(x_i+x_{i+1}\right)\left(x_i y_{i+1}+x_{i+1} y_i\right)}/\left(6\alpha \right) \label{x_cen}, \\
	y_c &=&\sum{\left(y_i+y_{i+1}\right)\left(x_i y_{i+1}+x_{i+1} y_i\right)}/\left(6\alpha\right) \label{y_cen}, \\
	\alpha &=& \sum{\left(x_i y_{i+1}+x_{i+1} y_i\right)}/2 \label{alpha}
\label{centroids_def}
\end{eqnarray}
where the sums are between all neighboring pairs of vertices on a bubble. An example of the large scale point pattern and a zoomed in version showing the points inside the bubbles they represent are shown in Fig.~\ref{PointRow}(a) and (e), respectively.

To understand the nature of the disorder in the location of bubble centroids, we compare with different disordered point patterns. Three types of patterns with varying degrees of uniformity are analyzed. The first type is an ``Einstein pattern"; these  consist of points initially placed on a square lattice and then randomly displaced by kick sizes drawn from a Gaussian distribution. Varying the root mean square (RMS) displacements of the particles will tune the disorder in the patterns \cite{ATCpixel}. For the purposes of this study the Gaussian kicks come from a distribution whose standard deviation is $\delta=0.26b$, where $b$ is the lattice spacing. We choose this value to make the number variance for the Einstein patterns the same as the number variance for the ``Halton patterns" which are the second type of pattern analyzed. Halton patterns use points from a low discrepancy sequence \cite{HaltonSeq}. They are of interest because although they are non-crystalline they fill space quite evenly; these properties make them and other low discrepancy patterns favorable for use in {\it e.g.}\ Monte Carlo integration \cite{HaltonMonteCarlo,NiederreiterBook,KocisWhitenACM}. Making a Halton pattern in two dimensions is done by choosing two integers $\{j_1,j_2\}$ whose only common denominator is 1; each number is an independent seeding element for a list of numbers and our patterns have $j_1=2$ and $j_2=3$. The ${n^{th}}$ number in the sequence is determined by converting $n$ into a number with base $j_k$, writing the number in reverse order after a decimal point and converting this fraction back into base 10 representation. This is done for both seeding elements and the pair of numbers creates one point in the Halton pattern. The fourth cellular pattern is a ``Poisson" pattern where uncorrelated points are laid down by drawing numbers from a random number generator. Fig.~\ref{PointRow}(b-d) shows the sample point patterns. 

However, bubbles are not simply points but are actually highly polydisperse cells of a larger space filling pattern. Therefore we also study how the areas of bubbles are distributed throughout space. For this analysis to keep with the definitions for $\chi(q)$ and $h(D)$ from Eqs.~\ref{chidef} and \ref{hdef}, the bubble centroids are given a weight equal to the area of the bubble they occupy. To find the areas, the bubbles are first treated like a polygon and the polygonal area is calculated using Eq.~(\ref{alpha}). The curved edges of the bubbles are not accounted for in this initial calculation. Accounting for them makes the final calculation of the bubble area the polygonal area plus or minus the area under each of the circular arcs if the arc bends away or towards from the centroid of the bubble, respectively. The foams are space filling and have a packing fraction of $\phi=1$.

Similar to the point pattern analysis, we want to compare data from bubbles to data from other cellular structures. In simulation we are free to partition space however we choose as long as we maintain a packing fraction $\phi=1$; for this study we create cellular patterns from Voronoi constructions around the three types of simulated point patters described earlier in this section. A Voronoi construction tiles space by separating points into cells whose edges are lines equidistant from the two points that share that edge. Voronoi patterns are generated using an intrinsic MATLAB function. This function also identifies the locations of the vertices for each cell and all cells are polygons; therefore Eq.~(\ref{alpha}) is used once again to calculate the cell area. Voronoi constructions, especially those made around Poisson patterns, have been studied extensively but much of the work is beyond the scope of this paper \cite{OkabeSpatial2009}; here they are used to study the structure of cellular patterns around point patterns of known disorder and compare that to the structure of quasi-2d foams which are cellular patterns around point patterns of unknown disorder. Recently work on cellular patterns in the context of hyperuniformity by partitioning space using several methods including Voronoi constructions was published \cite{TorquatoCell,KimCellScaling2019}; it does not include any experimental data  nor does it consider the hyperuniformity disorder length.


\section{Results}

Using the methods described above we reconstruct three snapshots of the same foam as it coarsens. They are taken $\{6,10,18\}$ hours after its initial preparation and have $N=\{2767,1842,1099\}$ bubbles, respectively. The total number of bubbles decreases as the foam ages because foam coarsening involves small bubbles shrinking and large bubbles growing due to differences in Laplace pressure until eventually the small bubbles disappear. There is not only an overall decrease in the total number of bubbles but also an overall increase in both the mean bubble area $\overline{a}$ and the $\phi$-weighted average bubble area $\left< a \right>$. Individual foam data sets are referred to by their value of $\left< a \right>=\{10,15,25\}~\text{mm}^2$; for polydisperse systems like the bubbles and Voronoi cells the $\left< a \right>$ is calculated by $\left< a \right>=\sum{{a_i}^2}/\sum{{a_i}}$ where the $a_i$ are the bubble or cell areas and the sum is over all particles. For the simulated data the Voronoi constructions are made in a square box bounded by $\left(0,1\right)$ with $N \geq 4.97 \times 10^5$ cells each. Only one Voronoi construction is generated for each type of point pattern.

\subsection{Local Properties}

\begin{figure}[t]
\includegraphics[width=3in]{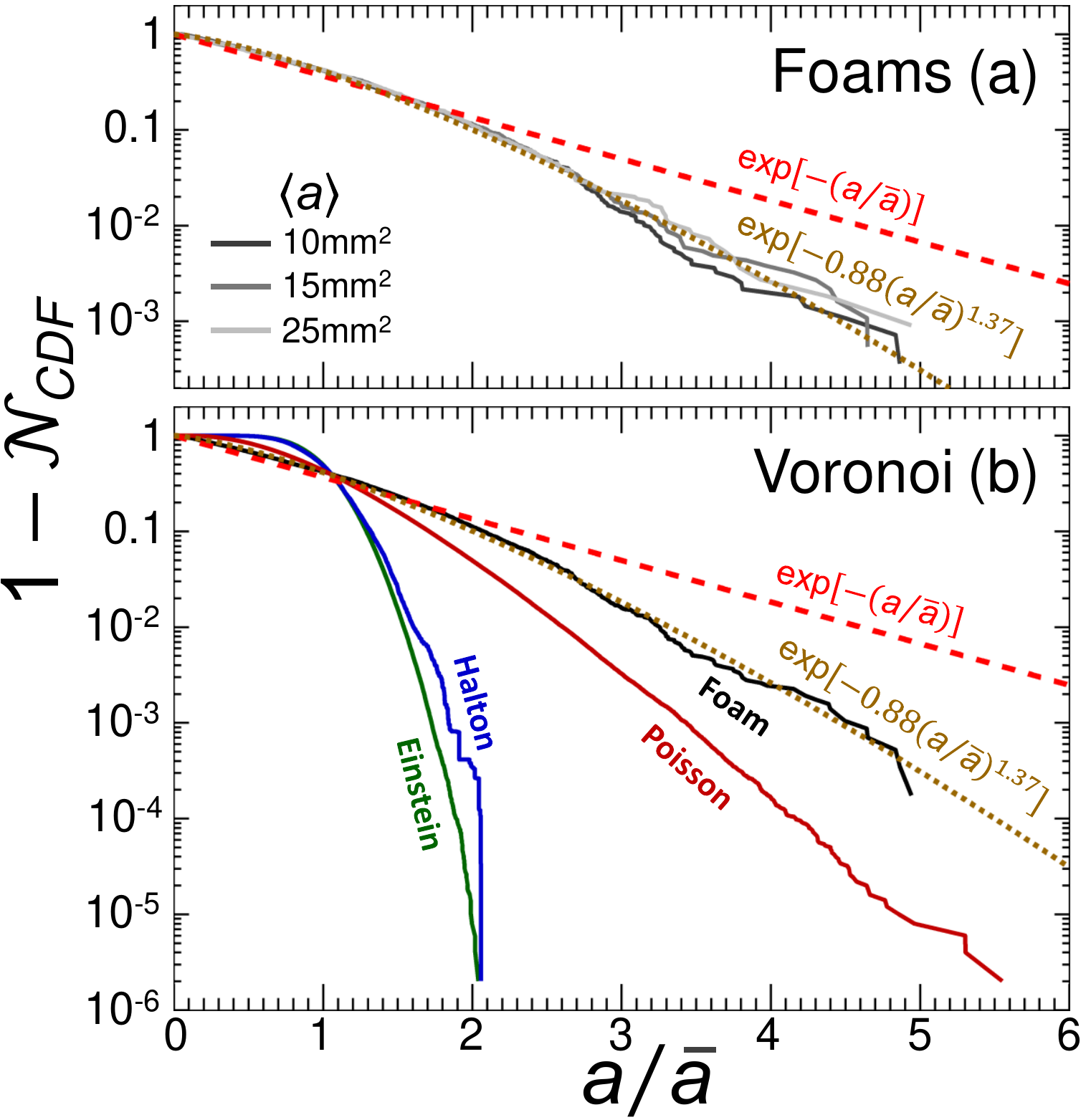}
\caption{Cumulative distribution function data for (a) bubble areas for foam as it coarsens and (b) areas of Voronoi cells constructed around point patterns as labeled. In part (a) the bubble areas collapse after normalizing by the mean area $\overline{a}$. In part (b) all the foam data are collected into one distribution and plotted as the black curve. In both parts the red dashed line shows an exponential area distribution and the gold dotted curve is a compressed exponential.}
\label{AreaCDF}
\end{figure}

Though it's not our main interest, for orientation and completeness we start by investigating several usual local structural features, beginning with the distribution of bubble areas. Fig.~\ref{AreaCDF}(a) shows the cumulative distribution function for the bubble areas normalized by the mean bubble area for the three snapshots of the coarsening foam. Amazingly, the data collapse regardless of the foam age. This is in fact a phenomena for foams aging referred to as a self-similar scaling state where distributions of local quantities are unchanged under proper normalization regardless of the age of the foam. Statistically, older foams are the same as taking a smaller subsection of a younger foam. Self-similarity is well documented and has been observed in experiment  \cite{StavansPRA,StavansPhysicaA, StavansGlazierPRL,deIcazaJAP76} and simulation \cite{GlazieretalPMB,HerdtleArefJFM,SegelEtAlPRE,RutenbergMcCurdyPRE,NeubertSchreckPhysicaA}. It is once again found here and the data are fit well to a compressed exponential consistent with previous work \cite{GlazierWeaireRev,StavansRev,RothPRE2013}. 

In addition to providing insight into the local structure of the foam the collapse of these distributions serves two more purposes. First it shows our methods for calculating the bubble areas are correct, which is very important for our hyperuniformity analysis. Second because the foam is in a scaling state the data from the three images can be collected together to make one distribution with better statistics. This is done for the normalized bubble areas and the data is plotted in Fig.~\ref{AreaCDF}(b) as a black curve. Comparing the cumulative distributions of cell areas for the Voronoi constructions to the bubble area distribution shows the latter is the widest. This demonstrates the local structure of the foam is the most disordered. The distributions for cell areas from the Voronoi constructions show the cells generated around the Einstein and Halton patterns have the most local order with nearly identical distributions and the cells generated around the Poisson patterns have a local order between Einstein/Halton and the foams. The local disorder is thus quantified by the width of the area distributions; one way to do this is by taking the mean squared cell area $\overline{a^2}=\sum{{a_i}^2} / N$ and dividing it by the mean cell area squared $\overline{a}^2=\left(\sum{{a_i}} / N\right)^2$; we find $\overline{a^2} / \overline{a}^2$ for each distribution in Fig.~\ref{AreaCDF}(b) and present their values in Table~\ref{area_distribution_table}. 

\begin{figure}[t]
\includegraphics[width=3in]{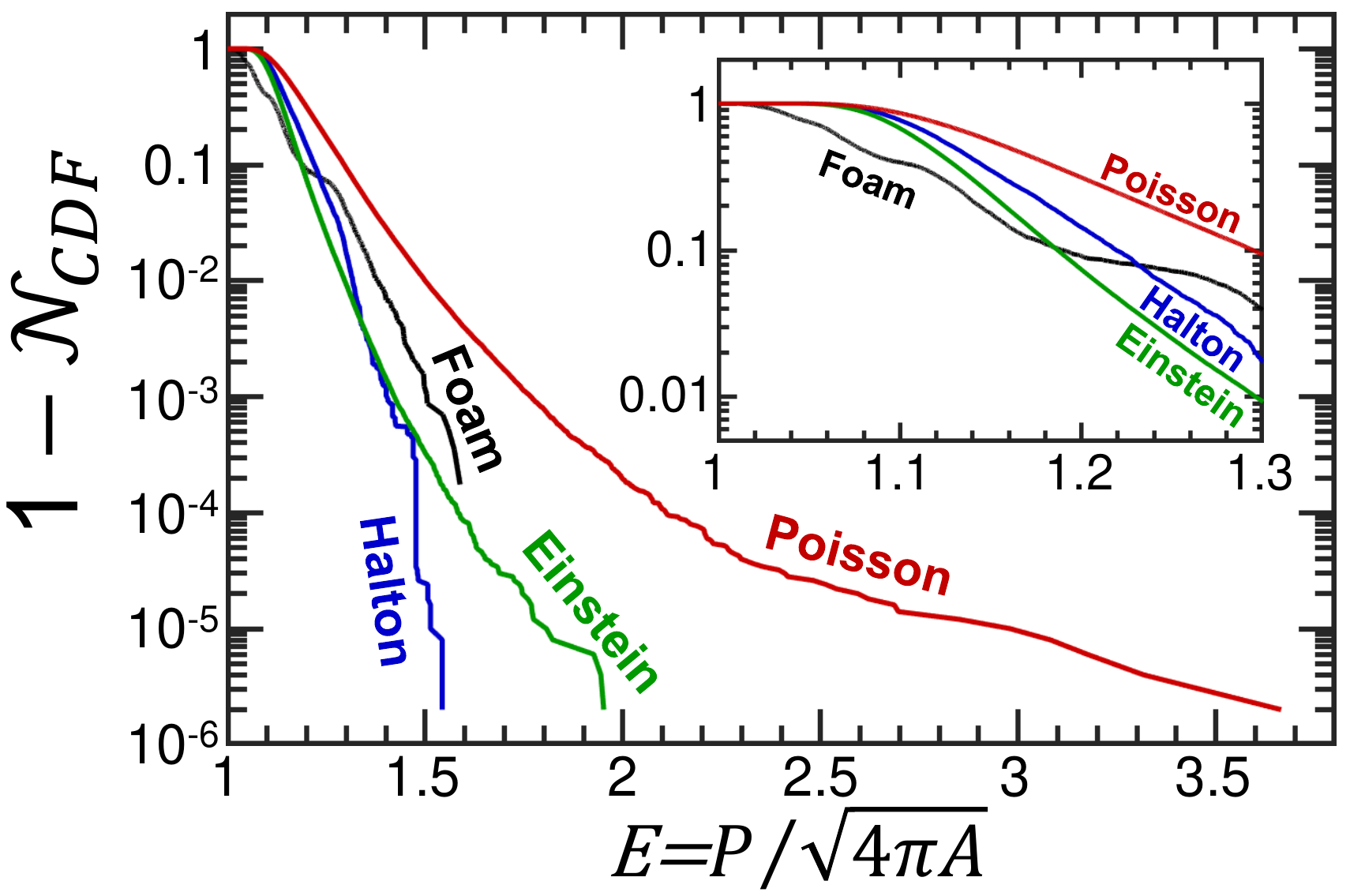}
\caption{Distributions of the elongation shape parameter for the various cellular patterns as labeled. The foam data is collected from the combined data from the three different times during the aging process. The distributions have statistical uncertainties described in Ref.~\cite{RothPRE2013} but the error bars are smaller than the symbol.}
\label{ElongCDF}
\end{figure}

The area is made dimensionless by dividing out the average area of at time $t$ but we can quantify other dimensionless shape parameters. One such parameter is the ``elongation" $E=P/\sqrt{4 \pi A}$ which takes the ratio of the bubble or cell perimeter to the square root of its area and it is defined such that $E=1$ for circles. Ref.~\cite{RothPRE2013} finds elongation to be one of two dimensionless shape parameters important in the physics of foam coarsening with the other being ``circularity"; circularity is defined to be 0 for polygons so we can not compare the quantity between the bubbles in a foam and the cells in a Voronoi construction. Calculating the elongation for all bubbles and collecting them into one distribution we compare the data to the elongation of the Voronoi cells. The distributions are shown in Fig.~\ref{ElongCDF} and the inset of the figure is a zoom in for the small $E$ data. The distribution for the foam is not the widest like it was for the areas but instead is smaller than the Poisson and goes further than both Halton and Einstein. We show the average elongation and the average squares elongations in Table~\ref{area_distribution_table} and in both cases these values are ordered from low to high as foam, Einstein, Halton and Poisson. Foam has the smallest average values because the data plunge away from $1$ the fastest which is seen clearly in the inset of Fig.~\ref{ElongCDF}. 

\begin{figure}[t]
\includegraphics[width=3in]{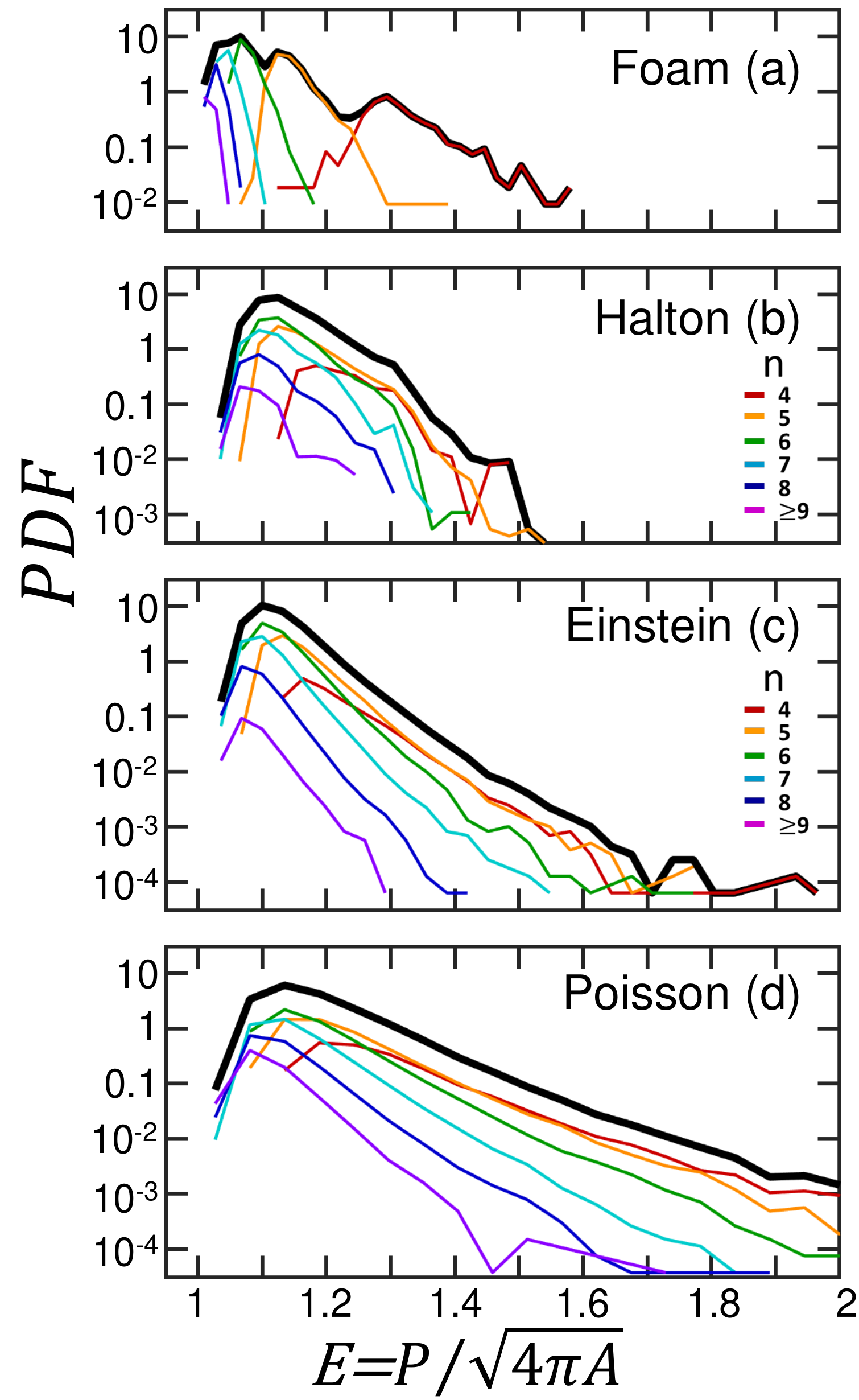}
\caption{Distributions of the elongation shape parameter for different packings as labeled. The actual distribution is plotted as a black line and data for $n$-sided cells are colored according to the number of sides. The foam data is collected from the combined data from the three different times during the aging process. }
\label{ElongPDF_sides}
\end{figure}

It is generally true for foams that bubbles with less sides have smaller areas and, similarly, we can ask how the number of sides affects the elongation. This is plotted in Fig.~\ref{ElongPDF_sides} where each part shows the elongation distribution for the entire packing along with the individual contributions to the distribution for $n$-sided cells. Fig.~\ref{ElongPDF_sides}(a) shows the data for the foam where data with small $E$ have large $n$ and bubbles with a smaller number of sides have larger $E$ values. Interestingly the foam have regions with little to no overlap for different $n$-sided bubbles; this is exhibited by the peaks of the individual $n$-sided distributions nearly matching the entire distribution especially for bubbles with less than 7-sides. For the Voronoi packings these regions of little overlap do not exist and the peaks of the distributions are not separated. Only the foams have well separated elongation distributions for different $n$-sided cells.

\bgroup
\def\arraystretch{1.25}%
\begin{table*}[t]
\setlength{\tabcolsep}{10pt}
\begin{center}
\begin{tabular}{c c c c c c c} 
\hline
\hline
Pattern & $\overline{a^2} / \overline{a}^2$ & $\overline{n}$ & $\sigma_n$ & $\overline{E}$ & $\overline{E^2}$ & $\overline{s^2} / \overline{s}^2$ \\ 
\hline
Einstein  & 1.05 $\pm$ 0.002 & 5.99 $\pm$ 0.01 & 0.993 $\pm$ 0.002 & 1.127 $\pm$ 0.0001 & 1.271 $\pm$ 0.002 & 1.29 $\pm$ 0.03  \\
Halton  & 1.06 $\pm$ 0.002 & 5.99 $\pm$ 0.01 & 1.092 $\pm$ 0.002 & 1.142 $\pm$ 0.0001 & 1.308 $\pm$ 0.002 & 1.33 $\pm$ 0.04  \\
Poisson  & 1.28 $\pm$ 0.006 & 5.99 $\pm$ 0.01 & 1.332 $\pm$ 0.002 & 1.181 $\pm$ 0.0002 & 1.403 $\pm$ 0.003 & 1.42 $\pm$ 0.04  \\
Foam  & 1.82 $\pm$ 0.07 & 5.98 $\pm$ 0.08 & 1.17 $\pm$ 0.02 & 1.10 $\pm$ 0.02 & 1.22 $\pm$ 0.03 & 1.19 $\pm$ 0.01  \\
\hline
\hline
\end{tabular}
\end{center}
\caption{Quantities characterizing distributions for the cellular patterns. The columns are the cellular pattern type, the average squared area divided by the average area squared, the average number of sides of a cell, the standard deviation for the side number distribution, the average elongation, the average squared elongation, and the average squared edge length divided by the average edge length squared. Data for all bubbles at the three times are collected into one distribution because the foam is in a self-similar state.}
\label{area_distribution_table}
\end{table*}
\egroup

\begin{figure}[t]
\includegraphics[width=3in]{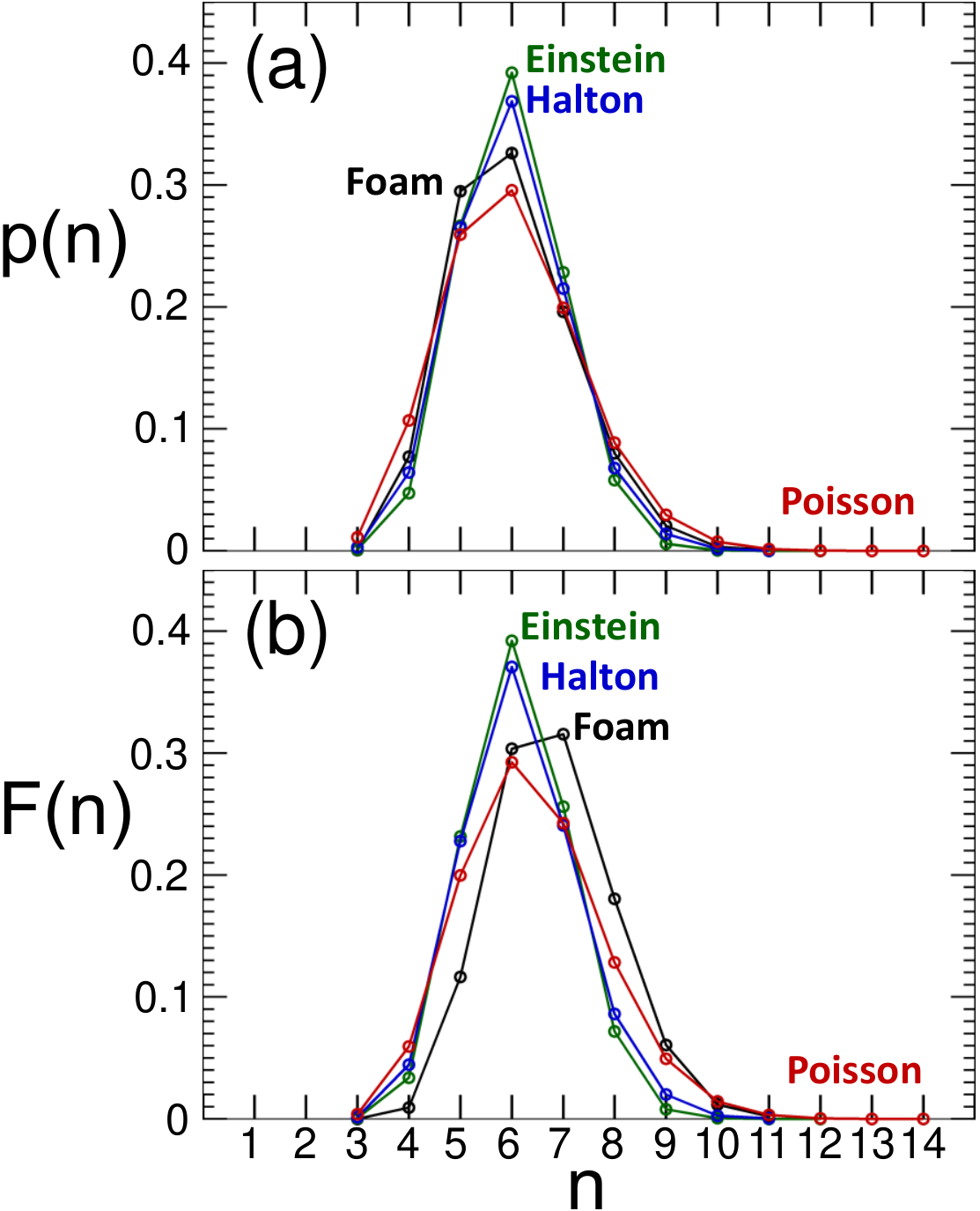}
\caption{(a) Side-number distributions, and (b) area-weighted side-number distributions, for the various cellular patterns as labeled. The foam data is collected from the combined data from the three different times during the aging process. The distributions have statistical uncertainties described in Ref.~\cite{RothPRE2013} but the error bars are smaller than the symbol.}
\label{pFofn}
\end{figure}

Other standard distributions we study include the side-number distribution $p(n)$  which tells the probability of finding a bubble or cell with $n$-sides and the area-weighted side-number distribution $F(n)$ which details how much area is covered by $n$-sided bubbles or cells. The distributions for $p\left(n\right)$ and $F\left(n\right)$ are plotted in Fig.~\ref{pFofn} parts (a) and (b), respectively. The $p\left(n\right)$ distributions are remarkably similar which is expected given that both the bubbles and Voronoi cells are convex polyhedra where the vertices are a junction of three edges; this microstructure also makes it so the average number of sides per cell is $\overline{n}=6$ and Table~\ref{area_distribution_table} shows this is almost exactly achieved for all packings. Part (b) shows the $F\left(n\right)$ distribution for the foam is skewed more towards cells with large $n$ when compared to the other distributions particularly for bubbles with $n=\left[7,8\right]$ sides. This is understood because bubbles with a larger number of sides also have larger areas. These distributions allow us to understand the local structure of the cellular patterns but next we investigate whether they provide any insight into the long range structure.

\subsection{Spatial Fluctuations of Number Density}

This and the two remaining subsections contain our main results, which concern the nature of long-range fluctuations in space-filling cellular structures.  We begin with number density fluctuations for all of the point patterns. For this analysis the points are all given an equal weight $w=1$ and distances are normalized by the square root of the $\phi$-weighted average area $\sqrt{\left<a\right>}$.

\begin{figure}[t]
\includegraphics[width=3.25in]{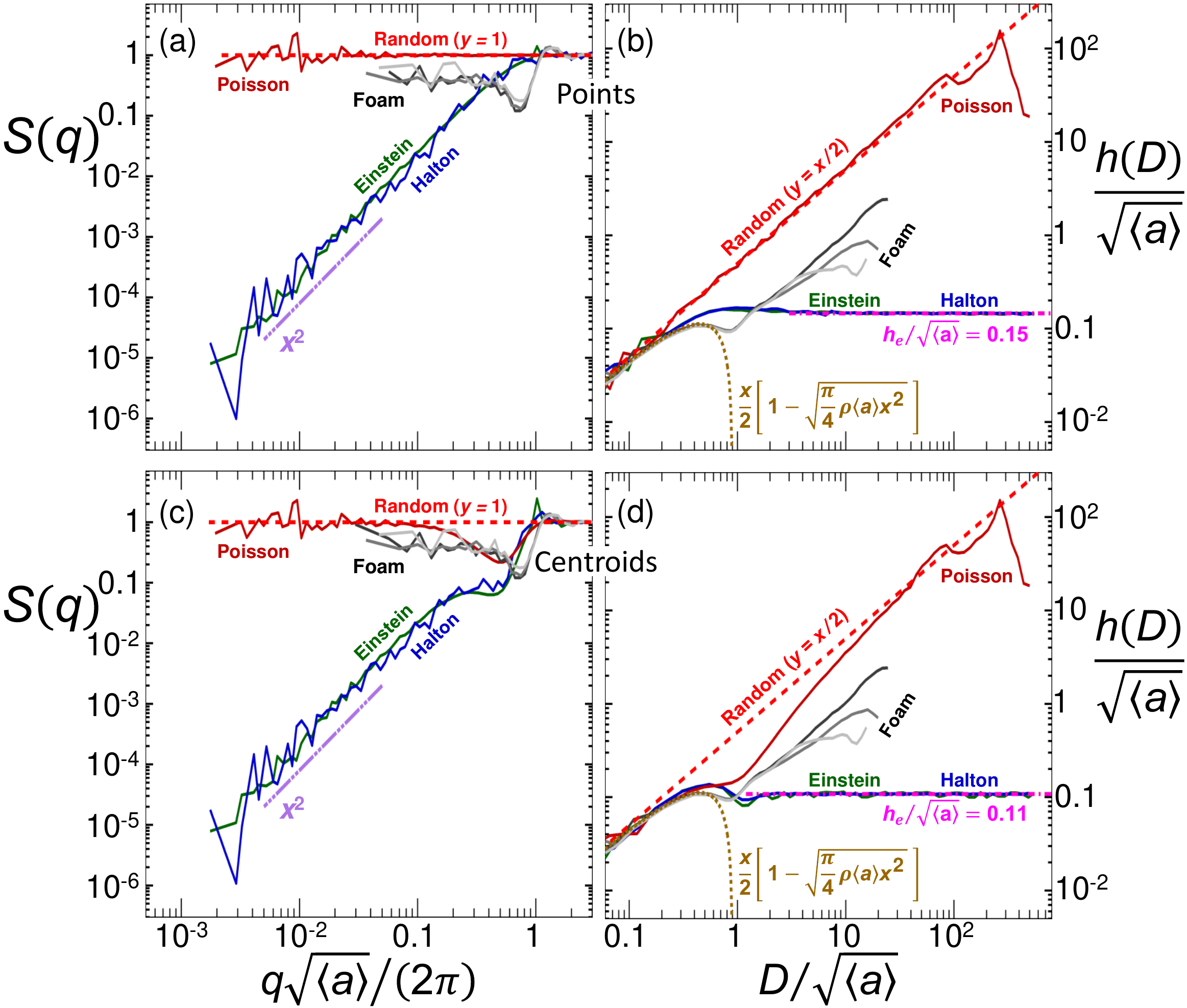}
\caption{Structure factor and associated real-space hyperuniformity disorder length spectra for various point and centroid patterns as labeled. The foam data, for systems with $\left<a\right>=\{10,15,25\}~\text{mm}^2$ as the curves go from dark to light gray, are the same in either parts (a)/(c) or (b)/(d) and have long range Poissonian fluctuations. The data for the simulated point patterns are as follows: the Poisson data lie along the random expectation (red dashed line); the Halton/Einstein data are hyperuniform indicated by $h(D)=h_e$ (magenta dot-dashed line) and by the power law decay of $S\left(0^+\right) \sim q^2$ (purple double dotted-dashed line). The centroid data are the same as the points data at long distances but there is an induced short range order at short distances; this additional order is continued to long distances for the Einstein/Halton centroids indicated by a smaller value of $h_e$.}
\label{PointsDataComp}
\end{figure}

The two ways we diagnose hyperuniformity in our point patterns is with the small-$q$ scaling of the structure factor $S(q)$ and with the large-$D$ behavior of the hyperuniformity disorder length. In Figs.~\ref{PointsDataComp}(a,b) the structure factor and hyperuniformity disorder length are plotted, respectively. For both quantities it is clear that data for the Poisson point patterns follow the totally random expectation. This is important for two reasons. The first is it confirms our analysis tools are working correctly for both the structure factor and the hyperuniformity disorder length. That is because the points in the Poisson pattern are totally uncorrelated and should follow Poisson statistics which they do. The second is that values of both $S(q)$ and $h(D)$ are made meaningful because the order is determined by how much smaller the data are than the upper bound set by the Poisson limit.

Poisson patterns are an example of completely random systems. Conversely, Einstein patterns are examples of systems we know are hyperuniform and previous work shows their uniformity is linked to $\delta$, the size of the RMS displacement of the particles away from their lattice site \cite{ATCpixel}. Hyperuniformity in the Einstein patterns is evident from the asymptotic behavior of $S(q) \sim q^2$ for small $q$ and $h(D)=h_e=0.15 \sqrt{\left< a \right>}$ for large-$D$. Recall these Einstein patterns have a $\delta=0.26b$ where $b$ is the lattice spacing so $h_e=0.15 \sqrt{\left< a \right>} \approx 0.55\delta$; this is consistent with previous work and so is the decay exponent for the structure factor \cite{ATCpixel,ATCjam}. 

We had no {\it a priori} knowledge about the long range order in the Halton pattern but the data show they too are hyperuniform. Their data behave nearly identically as the Einstein patterns but this similarity is no accident. Recall in Sec.~\ref{methods} that the kick size $\delta=0.26b$ was chosen such that the measured number variance for the Einstein patterns and Halton patterns are the same. In actuality the value of $\delta$ was determined by varying it until one was found to make the value for $h_e$ for the Einstein pattern match the value of $h_e$ of the Halton pattern. What is rather amazing here is that we matched the long range disorder of the point patterns and from that the distributions of Voronoi cell areas and topology are nearly identical. We have a direct observation at least for our example systems how the microscopics (locations of points around which Voronoi cells are constructed) affects the macroscopics (distributions of cell areas and topology).

For the foams each snapshot is analyzed separately and in  Fig.~\ref{PointsDataComp} plot the individual data sets for $\left<a\right>=\{10,15,25\}~\text{mm}^2$ as the curves that go from dark to light gray. The structure factor for the bubble centroids is interpreted as follows: at large $q$ the bubbles are initially random; there is short range order at approximately the average bubble separation indicated by a decay away from $\chi \left( q\right)=1$; for small $q$ there Poissonian fluctuations indicated by a leveling off to a constant. This behavior is mirrored in the hyperuniformity disorder length spectra. The data initially follow the separated particle limit for small $D$ because the bubble centroids have a minimum point separation based on average bubble size. The $h(D)$  spectra follow this expectation until $D \approx \sqrt{\left< a \right>}$ where the data reach some local minima but very quickly rise with $h(D) \sim D$ indicating Poissonian number density fluctuations. Rather remarkably the data collapse for the three ages of foam in both real and $q$-space; we interpret this to mean the arrangement of the bubble centroids while uncorrelated at long lengths does not change on average as the foam coarsens. This is likely an additional signature of the self-similar state of foams but now one that is observed at long distances.

The foam point patterns are unique among the types of point patterns we study because in Fig.~\ref{PointsDataComp}(a-b) they are the only pattern where the points lie at the centroid of their cell. Thus far we have only analyzed our three types of point patterns. However we used those patterns to seed Voronoi constructions so each point exists within a Voronoi cell; to make more direct comparisons to the foam data we make three new point patterns where we find the centroids of the Voronoi cells. The exact same foam data is plotted in Fig.~\ref{PointsDataComp}(c),(d) as in parts (a),(b) and compare it to the data for these ``centroid patterns"; for simplicity we will continue to refer to the data of these centroid patterns by the Voronoi construction seeding patterns. This naming convention is justified in Fig.~\ref{PointsDataComp}(c) because the data at small-$q$ for the centroid patterns the same as the data for the initial point patterns. The only difference in the $S(q)$ data for the centroid patterns is at intermediate-$q$ they have an initial dip, similar to the bubbles data, from an imposed length scale due to average size of the Voronoi cells. However the fact that the data is nearly identical at small-$q$ indicates the centroid patterns have some memory of the initial patterns used to seed the Voronoi cells. 

There is a similar effect for the hyperuniformity disorder lengths shown Fig.~\ref{PointsDataComp}(d). Here the induced short range order is exhibited because the $h(D)$ spectra are qualitatively the same as the separated particle limit; the curves do not match exactly because the expectation was plotted using the number density for the foam data. At large-$D$ the Poisson centroid data return to the random expectation. This ``memory" effect can be explained because the relative displacements from the seeding points to the centroids of the Voronoi cell creates a short range repulsion based on the average size of the cells and the points can not overlap. However points at long distances are still totally uncorrelated. Therefore fluctuations occur throughout the entire window for window sizes that are large compared to the distance between seeding points and the centroids of the Voronoi cell. 

No such memory exists for the hyperuniform patterns because the average particle displacement is relatively large compared to $h_e$. For the Einstein patterns, points have an RMS displacement of about $0.26$ the lattice space so the Voronoi cell constructed around them almost certainly has their initial lattice site in it. Creating the pattern for Voronoi cell centroids simply constructs a different Einstein pattern with a different kick size; the centroids have $h_e=0.11 \sqrt{\left< a \right>}$ and extracting a kick size from $h_e=0.55\delta$ finds a new smaller displacement is $\delta=0.2b$. This informs us that moving the points to the centroids of their Voronoi cell simply moves the seeding points closer on average to their original lattice site. Interestingly, the Halton centroid patterns continue to have the same long range number density fluctuations and the same value of $h_e$ as the Einstein patterns even after every point in both patterns is individually displaced. The fact that these centroid patterns are more ordered than the point patterns that are used to generate them but remain statistically identical is only seen by comparing the spectra of hyperuniformity disorder lengths; the structure factor have small-$q$ data that are the same for both the point and centroid patterns. For hyperuniform systems with small values of $h_e$, meaningful physical insight is gained into the spatial distribution of the points even with very small perturbations to their initial position. 

Finding number density fluctuations in point and centroid patterns opens up some interesting new avenues of research in particular in the case of scaling state foams and the general similarity between the Einstein and Halton patterns. However in the context of hyperuniformity the proper metric to study is the spectral density and fluctuations in area fraction for any pattern where particles have an area. This is also the analysis that is perhaps informed by the local distributions of bubble and cell sizes.

\subsection{Spatial Fluctuations of Area Fraction}

\begin{figure}[t]
\includegraphics[width=3.25in]{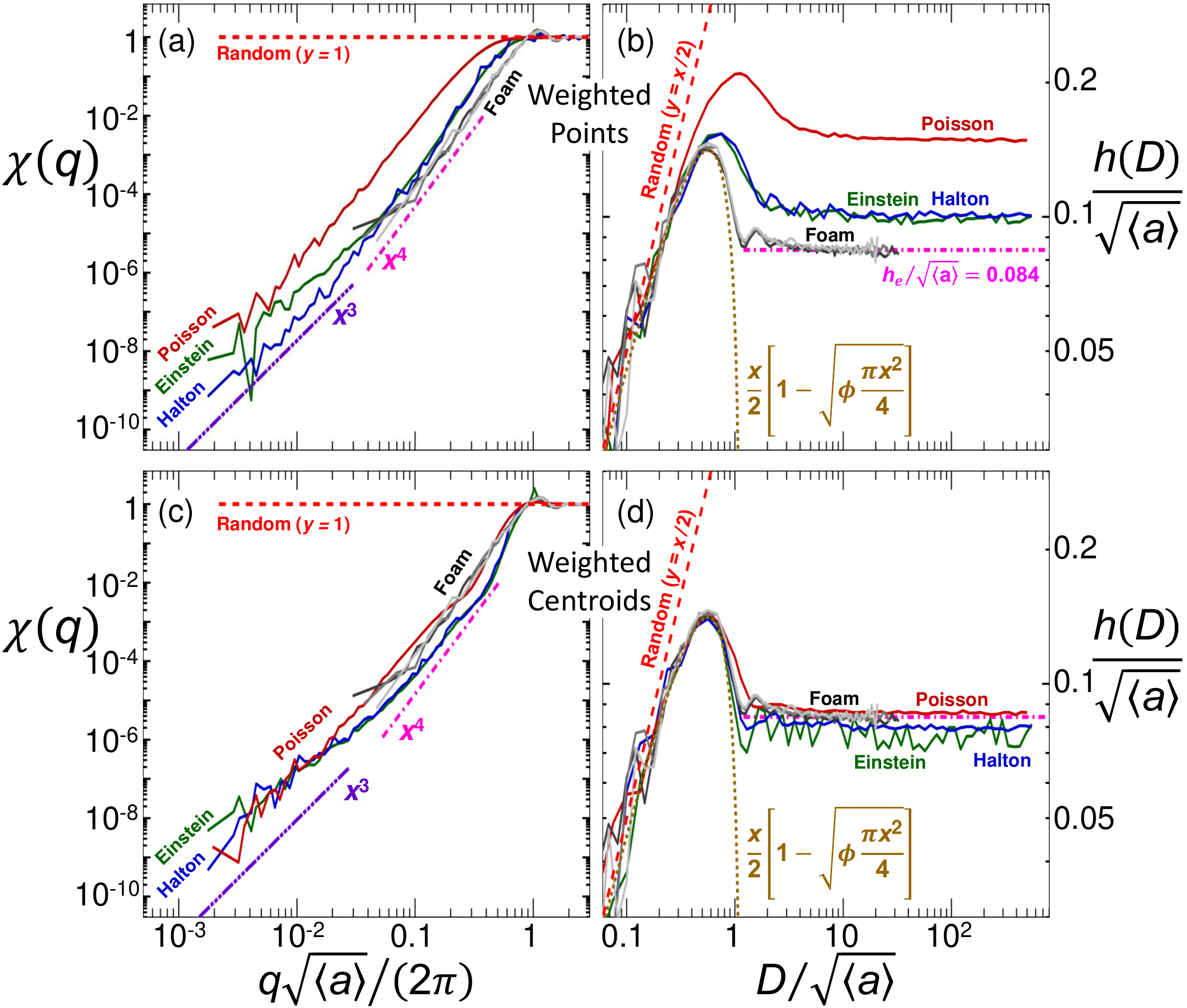}
\caption{Spectral density and associated hyperuniformity disorder length spectra for various cellular patterns as labeled. The foam data are for systems with $\left<a\right>=\{10,15,25\}~\text{mm}^2$ as the curves go from dark to light gray and are the same in either parts (a)/(c) or (b)/(d). In Part (b), the spectra of  $h(D)$ for all patterns at intermediate and long lengths becomes constant. The values of $h_e$ are different depending on the disorder and $h_e$ for the foam centroids is marked in both (b),(d) as a magenta dashed-dotted line. The spectral density decays like $q^{4.2}$ for the foam data for all $q\sqrt{\left< a \right>}/\left(2 \pi \right)<1$; for the Voronoi point and centroid data the $\chi\left(q\right)$  decay more slowly than the foam data but still have signatures of hyperuniformity. Only the foam data have a decay exponent near the $\epsilon=4$  expectation determined in \cite{KimCellScaling2019}. In part (d), the $h(D)$ nearly collapse following the separated particle limit (gold dotted curve) at small-$D$ and have very similar values of $h_e$.}
\label{PhiDataComp}
\end{figure}

In order to measure the spectral density and the hyperuniformity disorder length with regards to area fraction fluctuations each point is given a weight equal to the area of the cell it occupies in accordance with the central point representation. We start with the real space analysis. Because the patterns are space filling every observation window will be entirely covered in its interior and only the cells along the boundary will determine differences in packing fraction from $\phi=1$; this is essentially the definition of a hyperuniform system so we might expects all $\phi=1$ configurations with a reasonable size distribution of cells are trivially hyperuniform. This is borne out in Fig.~\ref{PhiDataComp}(b) where we convert the real space area fraction variance to a spectra of hyperuniformity disorder lengths for the various cellular patterns and all of the spectra have $h(D)=h_e$ for large measuring windows.

Unique to the spectra of hyperuniformity disorder lengths is that they not only determine whether a system is hyperuniform but also provide a meaningful length scale for the disorder. The value of $h_e$ indicates the distance from an observation window boundary where particles set the area fraction fluctuations; smaller $h_e$ means more order and we find the Poisson, Halton/Einstein, foam patterns are the least to most ordered. This lines up with our basic intuition for the Voronoi patterns: the Poisson point pattern is the most disordered and has the largest $h_e$; the unweighted data for the Einstein and Halton point patterns have matching long range order and so too does the area-weighted data. This one to one comparison of the unweighted point pattern data to the weighted point pattern data breaks down for the bubble centroids. The weighted bubble centroids data have area fraction fluctuations which are more suppressed at long lengths than the fluctuations for any of the area-weighted Voronoi point patterns. This is made even more surprising by the fact that the bubbles are the most disordered locally of the various cellular patterns. However the actual spatial arrangement of the bubbles is most ordered as dictated by foams having the smallest value of $h_e$. 

This arrangement for the bubble locations corresponds with the points being at the centroid of each cell which is not the case for the area-weighted Voronoi point patterns. When HUDLS is performed on the area-weighted Voronoi centroid patterns the data  in Fig.~\ref{PhiDataComp}(d) show the values of $h_e$ nearly collapse to the same value as $h_e$ for the foam centroids. Recent studies have found other metrics also collapse as they anneal Voronoi constructions with thousands of updates to the location of the point inside the Voronoi cell to the centroid of the cell \cite{KlattUniversal2019}; here $h(D)$ shows a collapse after just one step and it would be interesting to see if repeated annealing collapse the data even more towards the $h_e$ value of the foams. 

We compare the $h(D)$ data to the spectral density in Figs.~\ref{PhiDataComp}(a,c) and observe similar trends. For all curves nominal hyperuniformity is observed because the spectral density data decay like $\chi(q) \sim q^{-\epsilon}$ with $\epsilon > 1$.  In part~(a) the data for the weighted Voronoi points for the Einstein and Halton patterns initially decay with some exponent close to $\epsilon=4$ but have a final asymptotic scaling closer to $\epsilon=3$. The crossover from the initial to the final scaling finishes with less than one decade of data left so the actual value of $\epsilon$ is unreliable. No such crossover exists for the weighted Poisson points and for nearly all values of $q$ where the data decay they are fit well to $\chi(q) \sim q^{3.5}$. From the definition of $\chi(q)$ it is determined that weighted Halton and Einstein point patterns have more uniformity than the weighted Poisson points because the smaller the value of the spectral density the more order. The data for the weighted Voronoi centroid patterns shows a total collapse of the values of $\chi(q)$. Though the data do not fit well to a power law over more than one decade the final decay has $\epsilon \approx 3$.

Good estimates for these decay exponents are required not only because they act as a proxy for order but also because recent theoretical work provides an expectation for their value.  In Ref.~\cite{KimCellScaling2019} the authors find that if a fundamental cubic cell with periodic boundary conditions is tessellated into $N$ disjointed cells $\{C_1, .C_j,C_{j+1}.. ,C_N\}$ then the tessellations are all hyperuniform under some conditions. For two dimensional Voronoi constructions the conditions are as follows; all $C_j$ have a maximum side length much smaller than the total side length of the system; each $C_j$ is represented by a point or hard particle entirely within the Voronoi cell; each point or particle within $C_j$ has an assigned area $\psi |C_j|$ where $0<\psi<1$ and $|C_j|$ is the area of the $j^{th}$ Voronoi cell. Spectral density analysis for these tessellations show they are all hyperuniform with small wave vector scaling like $\chi(q) \sim q^{2}$ if the cells are represented by points away from the their centroid or $\chi(q) \sim q^{4}$ if the cells are represented by points at their centroid. 

None of our Voronoi patterns between the area weighted Voronoi points and the area weighted Voronoi centroids have spectral density with exponents that match the expectation in Ref.~\cite{KimCellScaling2019}. Instead the area weighted points all have $\chi (q)$ decay faster than expected and the area weighted centroids all have $\chi (q)$ decay slower than expected. The discrepancies may arise for two reasons: the first is we do not use periodic boundary conditions for our Voronoi constructions; the second is we use a $\psi=1$ for our analysis. These conditions may affect the decay exponents for the Voronoi constructions but amazingly analysis on the foam patterns show excellent agreement with the expectations.

Only one decade of $\chi(q)$ decays for the foams and the data only measure to a relatively large $q$ when compared to the Voronoi data. However, where the foam data decay they have $\epsilon = 4.2$. This in very nearly the value of $\epsilon=4$ expected for cellular patterns where the area weighting is assigned to the centroid of the cell from Ref.~\cite{KimCellScaling2019}; this is the first experimental work that we are aware of to confirm this expectation. Furthermore it is interesting to note that only the foams which are naturally occurring actually have this value while the simulated systems which are larger and more locally ordered do not match the expectation.
Additionally the foams exhibit the fastest decay when compared to all the other weighted point or centroid patterns indicating the highest long range order. Improving order in the Voronoi constructions likely involves repeating the process of moving the points to the centroids of their Voronoi cell and then making new Voronoi constructions. Upon many repetitions of this process the data will become more ordered and the exponent should fall more in line with the expectation. However in 2d the points will then start to crystallize. It would be interesting to devise a method to create non-crystalline patterns that can evolve in a way that their data matches the foam data.

In this section the results have been discussed through the lens of  hyperuniformity. We make this choice because the space filling materials show suppressed density fluctuations in both real and Fourier space and they have the same asymptotic behavior that is normally associated with hyperuniform materials. However it should be noted that while this behavior is nominally hyperuniform it may not truly be representative of the phenomena. This means the patterns made by the cellular points and centroids are likely not endowed with the certain special properties that are sometimes seen in hyperuniform materials like the having of complete photonic bandgaps \cite{FlorescuPNAS2009,ManPNAS2013}. Instead this behavior is due to the fact that $\phi=1$ for reasons either described earlier in this section for real space measurements or in \cite{KimCellScaling2019} for Fourier space measurements. Whether these systems are actually considered hyperuniform or not, the power of HUDLS as an analysis tool is still obvious because it is able to identify subtle differences in the underlying structure in the patterns generated within and by the cellular structures. If the cells are not truly hyperuniform because they are space filling then it is worth investigating whether hyperuniformity presents itself in the other much less dense phase of these patterns.

\subsection{Spatial Fluctuations of Cell Boundaries}

Until now all of the analysis focuses on the locations and areas of the bubbles and cells. For the foam this analyzes the location of the gas phase of the material. However, foam is a two-phase medium and consists of both gas and liquid phases and the liquid is contained in the surface Plateau borders, vertices, and films of the foam. For purposes of this study all of the liquid containing elements of the foam are referred to as the ``film network". This film network is also what constitutes the structure of the foams and makes the faces that separate bubbles; similarly the walls of the Voronoi constructions allow us to differentiate between cells. For simplicity we will use the term ``edges" to discuss both the foam films and Voronoi cell walls.

To quantify the spectrum of spatial fluctuations in the distribution of the liquid phase in foams, and to test for hyperuniformity, we need to define both the locations of the edges and their lengths. The foam films are arcs of circles that connect two vertices and the equation of the circle that defines each film is determined in the reconstructions; for a film with arc length $s=r \theta$ its location is defined as the point on the arc that bisects the angle theta. The Voronoi cell walls that connect two vertices whose locations are $(x_i,y_i)$ and $(x_j,y_j)$ have $s$ equal to the distance between the vertices and a location defined by the midpoint. The term area is used for the edges just for simplicity and is calculated with $a=t s$ . It is clear that the length $s$ for both the films and walls is important but the thickness $t$ is arbitrary. It is true that films in foam actually have some thickness but this value of $t$ is constant. Additionally the decoration theorem instructs that all of the liquid for a foam in the dry limit can be concentrated at the vertices with no effect to its structure \cite{BoltonWeaire1991}. The Voronoi cell walls should not have a thickness at all as they are lines. In both cases by assigning a constant value of $t$ to all the edges it drops out completely from Eq.~(\ref{chidef}) and while it is not as obvious mathematically the same is true for the $h(D)$ calculations. We performed auxiliary measurements changing the size $t$ and it does not affect our results; here we set $t$ equal to the film thickness $t=\ell$ so it has appropriate units.

\begin{figure}[t]
\includegraphics[width=2.2in]{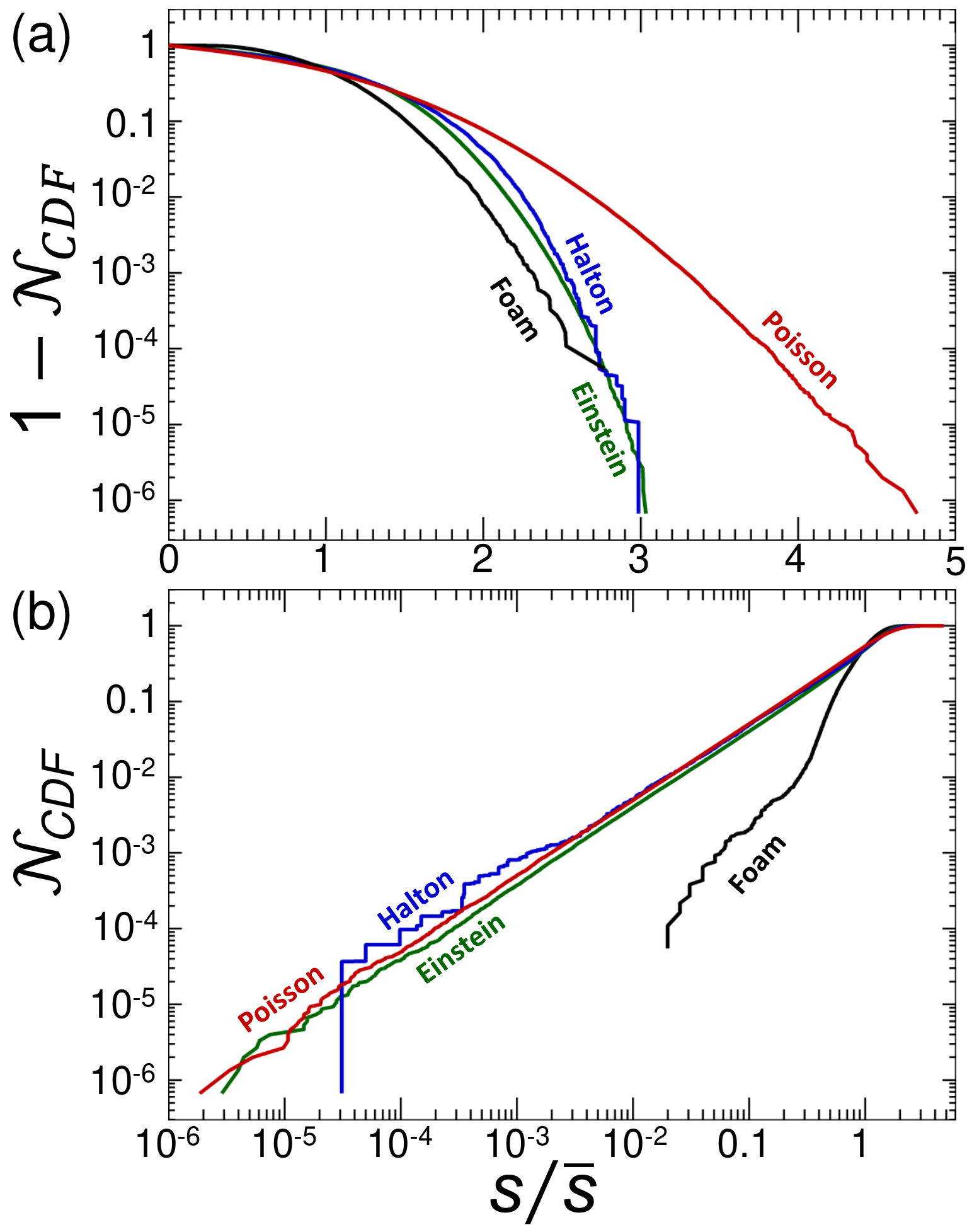}
\caption{The cumulative distribution function of edge lengths normalized by the mean edge length $\overline{s}$ for types of system as labeled. The edges for the foams (Voronoi cells) are the films (walls) that connect two neighboring vertices on a bubble (Voronoi cell) and they are circular arcs (straight line segments). In part (b) we show only the CDF on a log-log scale because the power law scaling where $\mathcal{N}_{CDF} \sim s / \overline{s}$ for the very small Voronoi wall lengths is lost when the CDF is subtracted from 1.}
\label{FilmLengthCDF}
\end{figure}

The only important feature then is $s$ the length of the edges. We measure values of $s$ for both the foam and Voronoi edges and plot them normalized by the mean edge length $\overline{s}$ in Fig.~\ref{FilmLengthCDF}.  Immediately evident is the foams have the narrowest distribution of edge lengths; this is juxtaposed with the fact that they have the broadest distribution of cell areas. This makes sense physically because Plateau's laws and coarsening both act to reduce the surface tension energy of the film network. No surface tension forces or size effects play a role in constraining the lengths of the Voronoi cell walls. The Poisson data is the widest which should be expected as there are large voids in these patterns which would lead to large edge lengths. The Einstein and Halton patterns have very similar distributions showing another macroscopic measure influenced by the microscopic point pattern. It is clear on the log-linear scale like Fig.~\ref{FilmLengthCDF}(a) the the lengths of the edges are relatively longer for the Voronoi networks than for the foams. The film length distributions are characterized similarly to the area distributions; here we use the mean squared edge length $\overline{s^2}=\sum{{s_i}^2} / N_s$ divided by the mean edge length squared $\overline{s}^2=\left(\sum{{s_i}} / N_s\right)^2$ where $N_s$ is the total number of edges and the results are displayed in Table~\ref{area_distribution_table}.

Not seen in Fig.~\ref{FilmLengthCDF}(a) is the fascinating behavior for small $s$. This is evident in Fig.~\ref{FilmLengthCDF}(b) which shows a log-log plot of just the CDF  and these distributions have a power law scaling where $\mathcal{N}_{CDF} \sim s / \overline{s}$ for all three types of the Voronoi constructions. No such scaling exists for the foam films. This power law scaling for the Voronoi edge lengths is rather remarkable and it is a result of the vertices of these Voronoi construction being overdispered compared to a Poisson patterns.

\begin{figure}[t]
\includegraphics[width=3.25in]{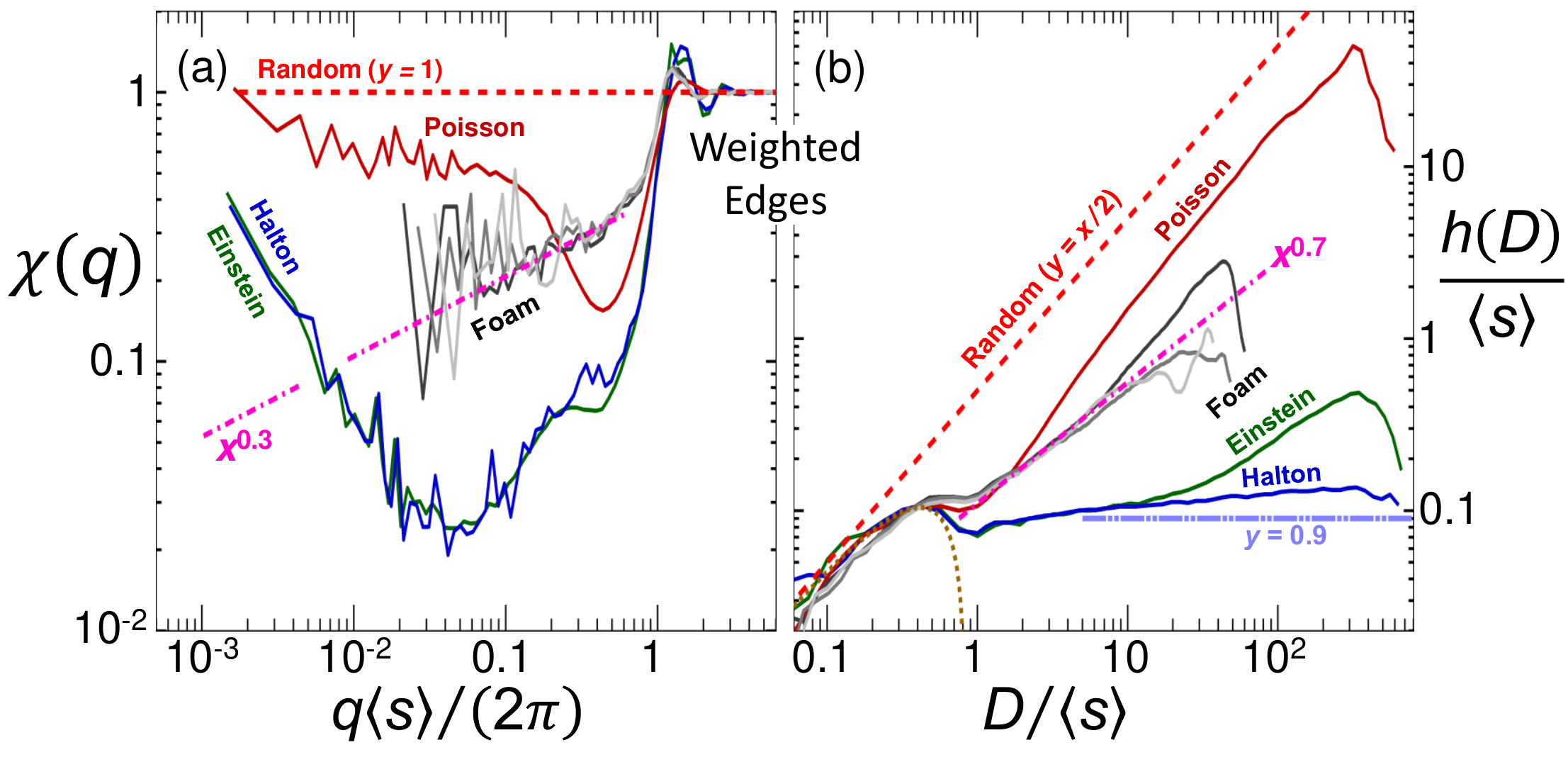}
\caption{Spectral density and associated hyperuniformity disorder length spectra for weighted edge patterns as labeled. In both parts the Poisson patterns show long range random fluctuations but do not lie exactly on the random expectation (red dashed line). The Einstein and Halton patterns are not hyperuniform; this is more evident in the spectral density data where the data have clearly defined minima but part (b) shows neither the Einstein nor the Halton data have $h(D)=h_e$ at large window sizes which is evident by comparing the data to a fiduciary constant (double-dot dashed line). The power law growth (magenta dot-dashed line) of $h(D) \sim D^{1- \epsilon}$ and $\chi(q) \sim q^{\epsilon}$ where $\epsilon=0.3$ is consistent with a class of hyperuniform materials.}
\label{WeightedFilmDataComp}
\end{figure}

We know from the previous section that local distributions do not always predict long range uniformity. In Fig.~\ref{WeightedFilmDataComp}(a) and (b) the data for the spectral density and the hyperuniformity disorder length with regards to area fraction fluctuations is plotted. To normalize the lengths in this figure the length scale $\left< s \right>= \sum{s^2} / \sum{s}$ is used; this is akin to our $\phi$-weighted average area. 

The spectral densities for the length-weighted Voronoi edge patterns shows that none of them are hyperuniform because each spectrum has some minimum before turning up towards the random limit. In real space the $h(D)$ for the length-weighted patterns confirm the Poisson edges are random because $h(D) \sim D$ for large $D$. However, there is ambiguity in the spectra of hyperuniformity disorder lengths for the Einstein and Halton edges. For these latter two cases there is less than a decade of data to potentially fit a power law to and the spectral density shows without a doubt that these systems are not hyperuniform. For the yes or no question of hyperuniformity in this case we defer to $\chi(q)$ and note that $h(D)$ does not equal a constant for either of these weighted Voronoi edge spectra. The values of $h(D)$ for the Halton data do lie below those for the Einstein data in a rare instance, but not unique see Fig.~\ref{PhiDataComp}(a), where the data are not practically the same. It is possible that the local structure informs this difference in uniformity and Fig.~\ref{FilmLengthCDF} shows why. The data show the Halton edges appear to have some minimum cutoff length scale and this is not the case for the Einstein edges. These very small-length edges in the Einstein patterns could form more dense clusters of edges which in turn makes them less uniform with regards to area fraction fluctuations; this merits further study and if true would be quite amazing that such a small fraction of points can destroy overall uniformity. 

Unlike for the Voronoi edges, the  spectral density for the length-weighted edges of the foam do not exhibit any minimum. The data are noisy and it is unclear whether they could be either decaying like a power law and hyperuniform or leveling off to a constant and Poissonian. A clearer signal comes from the hyperuniformity disorder length. The foam data is well fit to a power law over one decade of window sizes; the power law exponent $\epsilon=0.3$ shows long range uniformity in both real and Fourier space and follows the expectation that $h(D) \sim D^{(1- \epsilon)}$ and $\chi(q) \sim q^{\epsilon}$. This shows a weaker variant of hyperuniformity than if $h(D)$ were some constant but nonetheless fluctuations are suppressed at long length scales. The data for the weighted foam edges are not the most uniform in an absolute sense because for this to be true the spectra of $h(D)$ and $\chi(q)$ would have to have smallest values like they do for the weighted point data. However they are the only edge pattern that has a legitimate signature of hyperuniformity. 

\section{Conclusion} 

We have presented the use of a recently defined emergent length scale, the hyperuniformity disorder length, as a method to describe the structure of various cellular patterns at all length scales. We have shown that usual local descriptors of cellular patterns like the cell area or side number distribution do little to properly differentiate the underlying disorder in the cellular structure. Similarly the asymptotic scaling of the spectral density fails to differentiate order because the exponents are not always obvious even for very large systems. However, one important result from the Fourier space analysis arises in the foams having a spectral density that decays for all small-$q$ like $q^{4.2}$. This is an unambiguous result and the decay is faster for foam data than it is for the Voronoi point and centroid data. Furthermore the foams which are the only naturally occurring patterns we study are also the only structures that have a spectral density decay with an exponent near the $\epsilon=4$  expectation set forth in Ref.~\cite{KimCellScaling2019}. To more clearly compare order between the packings we use hyperuniformity disorder length spectroscopy; the spectra of values of $h(D)$ provide a physically significant description to the meaning of both number and area fraction fluctuations and has helped to discover big differences in uniformity based on subtle differences in structure. 

Some of these findings are most apparent when comparing the data of Einstein and Halton point patterns. We saw that by tuning the underlying microscopic disorder in the Einstein pattern to match the disorder in the Halton pattern that we can construct nearly identical macroscopic patterns in terms of particle area and topology. Both patterns are hyperuniform with the same values of $h_e$. Additionally while a massive amount of particle rearrangement occurs when we shift these two point patterns to their centroid patterns, both the Einstein and Halton patterns had an overall increase in order which was the same for both patterns. This is only understood because the values of $h_e$ dropped but to the same value for both the Einstein and Halton pattern. Being able to differentiate these small structural differences after a lot of particle motion may be useful in studying the reversible to irreversible transition; in these experiments particles can be tracked near the critical amplitude for the transition and the small differences in structure may be evident using HUDLS. 

Turning to foams we found them to be the most ordered of the area-weighted point patterns even though they are the most locally disordered. This is likely due their points being located at the centroid of the bubble as opposed to the point patterns weighted by the Voronoi cell areas that are located at the points that seeded the Voronoi construction. When we compare all the centroid patterns the the values of $h_e$ collapse nearly to the same value as the foams. This $h_e$ also happens to be the same value that soft disks above jamming have before an onset of Poissonian fluctuations \cite{ATCjam}. This value is potentially some universal minimum $h_e$ for disordered patterns; it is possible that if we continue to update the centroid patterns the $h_e$ will either converge to the value of $h_e$ for the foams or the systems crystallize and the $h(D)$ will have increasing oscillations. It is an open question as to whether 2d configurations can be constructed by updating Voronoi patterns and avoid crystallization. Foams are also the only system we study whose edges have any signature of hyperuniformity. This hyperuniformity is defined by the scaling exponent $\epsilon=0.3$ and  $h(D) \sim D^{1- \epsilon)}$ and $\chi(q) \sim q^{\epsilon}$. Also for foams we have found a long range signature of the scaling state with regards to number density fluctuations because both the hyperuniformity disorder length spectra and the structure factor are unchanged as the foam ages. It would be interesting to try and push this to system sizes on the order of $N=10^5$ like we did for the Voronoi constructions but this would require simulation. 

Besides foams we can use the hyperuniformity disorder length to determine long range disorder in other naturally occurring cellular patterns This analysis can be used in 2-dimensions on networks made from cracks in dried mud, from peaks and valleys in crumpled paper, or from biological cells. In 3-dimensions one could study biological networks of trabecular bone or any other types of porous materials. A natural extension of our work is to perform analysis for 3-dimensional foams. Recent experiments on 3d foam has found that they, like 2d foams, enter a self-similar scaling state \cite{LambertGranerPRL2010}. Applying HUDLS to any of these systems offers a general and intuitive real-space method to characterize the spectrum of structural features which is a fundamental step in understanding material properties.

\begin{acknowledgments}
We thank Nigel Goldenfeld for introducing us to the concept of low discrepancy sequences. This work was supported primarily by NASA grant 80NSSC19K0599 and also by NSF grant MRSEC/DMR-1720530.
\end{acknowledgments}

\bibliography{HyperRefs}
\end{document}